\documentstyle[12pt]{article}
\begin{document}

\title{Scheme for the generation of an entangled four-photon W-state}
\author{XuBo Zou, K. Pahlke and W. Mathis  \\
\\ Electromagnetic Theory Group at THT,\\
 Department of Electrical
Engineering, University of Hannover, Germany }
\date{}

\maketitle

\begin{abstract}
{\normalsize We present a scheme to produce an entangled
four-photon W-state by using linear optical elements. The
symmetrical setup of linear optical elements consists of four beam
splitters, four polarization beam splitters and four mirrors. A
photon EPR-pair and two single photons are required as the input
modes. The projection on the W-state can be made by a four-photon
coincidence measurement. Further, we show that by means of a
horizontally oriented polarizer in front of one detector the
W-state of three photons can be generated.

PACS number:03.65.Ud, 03.67.-a, 42.50.Dv}

\end{abstract}
Entanglement is one of the most striking features of quantum
mechanics. Entangled state of two or more particles not only
provides the possibilities to test quantum mechanics against local
hidden theory\cite{EPR, bell}, but also have practical
applications in quantum information processing, such as quantum
cryptography\cite{cry}, quantum dense coding \cite{qua} and
quantum teleportation\cite{tele}. Since the GHZ state was
introduced to test quantum nonlocality without
inequality\cite{ghz}, there has been much interest in the
investigation of multiparticle entangled state for more useful
applications\cite{ghz,kem,tha,gt}. In Ref\cite{duer}, it was shown
that there exist two inequivalent classes of three-qubits
entangled states, namely the GHZ class and the W class. These two
classes are inequivalent because states from one cannot be
obtained from the states of the other by local operation and
classical communication. Another key difference is that a
single-particle trace of the GHZ state result in a maximally mixed
state compared with a nonmaximally mixed result for the W
state\cite{duer}. In Ref\cite{wtele}, the W state has been used to
realize the teleportation of an unknown state probability.
Recently a different set of Bell inequality is considered with the
purpose of illustrating some difference between the violation of
local realism exhibited by the GHZ and W state\cite{wbell}. More
recently, it is shown that the correlation between two qubits
selected from a trio prepared in a W state violate the
Clauser-Horne-Shimony-Holt inequality more than the correlation
between two qubits in any quantum state\cite{bells}. In recent
experiment, complementary measurement also reveal difference
between the GHZ and W state\cite{hou}

Experiments with entangled photons open a broad field of research.
Entangled photon states can be used to test Bell's inequality
\cite{daa} or to implement quantum information protocols for
quantum teleportation \cite{db}, quantum dense coding \cite{km}
and quantum cryptography \cite{ds}. An experimental realization of
GHZ-states by means of three or four photons was experimentally
observed and used to verify quantum nonlocality\cite{bou,dj,pan}.
Following the scheme of efficient quantum computation with linear
optics\cite{erg}, a feasible linear optical scheme \cite{zxb} was
proposed to produce entangled $N$-photon states via linear optical
elements. The incentive to produce entangled states of spin-s
objects $(S>1/2)$ increased significantly in the context of
quantum communication and quantum information processing
\cite{gp,kp}. A scheme was proposed \cite{zxb1} to generate an
entangled four-photon state, which is equivalent to two maximally
entangled spin-1 particles. Drummond \cite{pdd} demonstrated the
violation of the Bell inequality by this kind of multi-photon
states. Experimental realization of spin-1 entangled state have
been reported\cite{pdd1}. In this paper, our aim is to propose a
scheme to generate a W-state
\begin{eqnarray}
|W\rangle=\frac{1}{2}(|HHHV\rangle+|HHVH\rangle+|HVHH\rangle+|VHHH\rangle)\label{1}
\end{eqnarray}
of four photons by means of linear optical elements, which is
realizable experimentally. We use an EPR-source $\Psi_{EPR}$ in
order to generate two entangled branches of the symmetric
experimental setup. These two branches are denoted by the mode $1$
and the mode $2$. Two single-photon sources are injected in order
to generate an entangled four-photon state. A four-photon
coincidence detection projects this state on the W-state. The
scheme is based on post-selection strategy, which has been used to
demonstrate quantum information processing and generate
multi-photon GHZ state\cite{db}-\cite{pan}

In the following we present a detailed analysis of the proposed
scheme. The experimental setup is shown in Figure 1. The
EPR-source generates the input state
$\Psi_{EPR}=(|H\rangle_1|V\rangle_2+|V\rangle_1|H\rangle_2)/\sqrt{2}$.
In order to inject two further photons in this arrangement of
linear optical devices we use two single-photon sources:
$\Psi_e=|H\rangle_e$ and $\Psi_f=|H\rangle_f$. These single-photon
input modes $e$ and $f$ are horizontally polarized. The
experimental setup required for the two-photon injection consists
of four polarization beam splitters $PBS_i$, two beam splitter
$BS_j$ and four mirrors. Since polarization beam splitter transmit
the horizontal polarization and reflect vertical polarization, the
two polarization beam splitters $PBS_1$ and $PBS_2$ transform the
EPR-state $\Psi_{EPR}$ into
\begin{eqnarray}
\Psi_1=\frac{1}{\sqrt{2}}(|H\rangle_b|V\rangle_c+|V\rangle_a|H\rangle_d)\,.
\label{2}
\end{eqnarray}
Notice, that the modes $b$ and $d$ pass through the beam splitters
$BS_{1}$ and $BS_2$, and that each of these beam splitters is
connected with one of the single-photon input modes. The result of
the two-photon injection at the beam splitters $BS_1$ and $BS_2$
is
\begin{eqnarray}
\Psi_2=\left(|2H\rangle_b|V\rangle_c|H\rangle_d+|V\rangle_a|H\rangle_b|2H\rangle_d\right)\cos\theta\sin^2\theta\nonumber\\
+\left(|H\rangle_b|V\rangle_c|H\rangle_e|H\rangle_f+|V\rangle_a|H\rangle_d|H\rangle_e|H\rangle_f\right)\cos2\theta\cos\theta/\sqrt{2}\nonumber\\
+\left(|H\rangle_b|V\rangle_c|H\rangle_d|H\rangle_e+|V\rangle_a|H\rangle_b|H\rangle_d|H\rangle_f\right)\cos2\theta\sin\theta/\sqrt{2}\nonumber\\
+\left(|V\rangle_c|2H\rangle_e|H\rangle_f+|V\rangle_a|H\rangle_e|2H\rangle_f\right.\nonumber\\
\left.-|2H\rangle_b|V\rangle_c|H\rangle_f
-|V\rangle_a|2H\rangle_d|H\rangle_e
\right)\cos^2\theta\sin\theta\nonumber\\
-\left(|V\rangle_c|H\rangle_d|2H\rangle_e+|V\rangle_a|H\rangle_b|2H\rangle_f\right)\cos\theta\sin^2\theta
\,.\label{3}
\end{eqnarray}
The reflectance $\sin\theta$ and the transmittance $\cos\theta$ of
these beam splitters are the same, since the same angle $\theta$
is chosen. This coefficient will be determined later. This
four-photon state interacts with the two polarization beam
splitters $PBS_3$ and $PBS_4$. The horizontal polarization is
transmitted and the vertical polarization is reflected. This
method of a two-photon injection on the EPR-modes $1$ and $2$
results in the four-photon state
\begin{eqnarray}
\Psi_3=\left(|2H\rangle_1|V\rangle_2|H\rangle_2+|V\rangle_1|H\rangle_1|2H\rangle_2\right)\cos\theta\sin^2\theta\nonumber\\
+\left(|H\rangle_1|V\rangle_2|H\rangle_e|H\rangle_f+|V\rangle_1|H\rangle_2|H\rangle_e|H\rangle_f\right)\cos2\theta\cos\theta/\sqrt{2}
\nonumber\\
+\left(|H\rangle_1|V\rangle_2|H\rangle_2|H\rangle_e+|V\rangle_1|H\rangle_1|H\rangle_2|H\rangle_f\right)\cos2\theta\sin\theta/\sqrt{2}
\nonumber\\
+\left(|V\rangle_2|2H\rangle_e|H\rangle_f+|V\rangle_1|H\rangle_e|2H\rangle_f\right)\cos^2\theta\sin\theta
\nonumber\\
-\left(|2H\rangle_1|V\rangle_2|H\rangle_f+|V\rangle_1|2H\rangle_2|H\rangle_e
\right)\cos^2\theta\sin\theta
\nonumber\\
-\left(|V\rangle_2|H\rangle_2|2H\rangle_e+|V\rangle_1|H\rangle_1|2H\rangle_f\right)\cos\theta\sin^2\theta)\,.
\label{4}
\end{eqnarray}
This expression shows the generated four-photon state explicitly
in terms of the input modes $(1,2,e,f)$. In order to perform a
projection on the W-state a four-photon coincidence detection is
needed. This can be done by splitting the branches $1$ and $2$
separately by the beam splitters $BS_3$ and $BS_4$.
\begin{eqnarray}
\Psi_4=\frac{\cos\theta\sin^2\theta}{2\sqrt{2}}[|H\rangle_1|H\rangle_3|H\rangle_2|V\rangle_4+
|H\rangle_1|H\rangle_3|V\rangle_2|H\rangle_4\nonumber\\
+|V\rangle_1|H\rangle_3|H\rangle_2|H\rangle_4+|H\rangle_1|V\rangle_3|H\rangle_2|H\rangle_4]
+\Psi_{other} \,.\label{5}
\end{eqnarray}
Since we consider only those terms, which correspond to a 4-photon
coincidence detection (one photon in each of the four beams), The
other terms $\Psi_{other}$ of this quantum state don't contribute
to the event which we are considering. Thus the state of the
system is projected with probability $\cos^2\theta\sin^4\theta/2$
into the W-state
\begin{eqnarray}
\Psi_5=\left(|H\rangle_1|H\rangle_3|H\rangle_2|V\rangle_4+|H\rangle_1|H\rangle_3|V\rangle_2|H\rangle_4\right)/2\nonumber\\
+\left(|V\rangle_1|H\rangle_3|H\rangle_2|H\rangle_4+|H\rangle_1|V\rangle_3|H\rangle_2|H\rangle_4\right)/2
\,.\label{6}
\end{eqnarray}
The probability attains it`s maximal value $2/27$, if the
transmittance of the beam splitter is $\cos\theta=1/\sqrt{3}$.\\
If a polarizer with horizontal orientation is used in front of
detector 1 the four-photon coincidence detection projects into the
three-photon state
\begin{eqnarray}
\Psi_6=(|H\rangle_3|H\rangle_2|V\rangle_4+|H\rangle_3|V\rangle_2|H\rangle_4
+|V\rangle_3|H\rangle_2|H\rangle_4)/\sqrt{3} \,.\label{7}
\end{eqnarray}
Now we consider the experimental realization of the present
scheme. Firstly we demonstrate that the requirement for two
single-photon sources and one two-photon EPR-state source can be
realized using technology presently. In the experiment\cite{dj},
two separate polarization entangled photon state are generated via
type-II down conversion pumped by a laser field. We will show that
a modification of the scheme illustrated in\cite{dj} can be used
to prepare two single-photon sources and one two-photon EPR-state
source. The output light of a pulsed laser with angular frequency
$\omega_0$ is divided into two by a beam splitter. Transmitted and
reflected parts of the light are frequency doubled in a nonlinear
crystal. Transmitted pulses of the frequency $2\omega_0$ is used
to pump a type-II phase-matched parametric down conversion crystal
arranged to emit polarization entangled photon state into the two
modes with different polarization shown in Ref\cite{kkk}.
Reflected pulses of the frequency $2\omega_0$ is used to pump a
type-I phase-matched parametric down conversion crystal arranged
to produce down converted photon pairs in to into the two modes
with same polarization shown in Ref\cite{mandel}, in which a two
photon interference effect has been observed. Another requirement
of the present scheme is four-photon coincidence detection. such
multi photon coincidence have been realized and used for the
preparation of multi-photon entangled state\cite{dj}. Therefore,
based on technology presently, the present scheme might be
realizable.

In conclusion, a scheme is presented to produce entangled W-states
of four photons based on linear optical devices and a four-photons
coincidence detection. Further on we proposed a modification of
this symmetric scheme by adding one horizontal oriented polarizer
in front of detector $D_1$. Such multi-photon states may play a
crucial role in fundamental tests of quantum mechanics versus
local realism and in many quantum information and quantum
computation schemes.

\begin{flushleft}

{\Large \bf Figure Captions}

\vspace{\baselineskip}

{\bf Figure 1.} The schematic of the W state of four photons. EPR
denote entangled two photon state
$\frac{1}{2}(|H\rangle_1|V\rangle_2+|V\rangle_1|H\rangle_2)$.
$PBS_i$ are polarization beam splitters and $BS_i$ are beam
splitters. $D_i$ are photon number detectors.
\end{flushleft}


\begin{thebibliography}{99}
\bibitem{EPR}A. Einstein, B. Podolsky and N. Rosen, Phys. Rev. 47, 777 (1935).
\bibitem{bell}J. S. Bell, Physics (Long Island City, N.Y.) 1, 195 (1965).
\bibitem{cry}A. K. Ekert , Phys. Rev. Lett. 67, 661 (1991).
\bibitem{qua}C.H. Bennett and S.J. Wiesner, Phys. Rev. Lett. 69, 2881 (1992)
\bibitem{tele}C. H. Bennett , G. Brassard, C. Crepeau, R. Jozsa , A. Peres, and W. Wootters , Phys. Rev. Lett. 70, 1895 (1993).
\bibitem{ca}
E. Hagley , X. Maitre, G. Nogues, C. Wunderlich , M. Brune, J. M.
Raimond , and S. Hroche, Phys. Rev. Lett. 79, 1 (1997).
\bibitem{tr}Q. A. Turchette , C. S. Wood, B. E. King , C. J. Myatt,
D. Leibfried , W. M. Itano, C. Monroe , and D. J. Wineland , Phys.
Rev. Lett. 81, 3631 (1998);M. A. Rowe , D. Kielpinski, V. Meyer,
C. A. Sackett, W. M. Itano, C. Monroe , and D. J. Wineland ,
Nature (London) 409, 791 (2001).
\bibitem{ghz}
D.M. Greenberger, M. Horne, A. Shimony, and A. Zeilinger, Am. J.
Phys. 58, 1131 (1990).
\bibitem{kem}
J. Kempe, Phys. Rev. A 60, 910 (1999).
\bibitem{tha}A.V. Thapliyal, Phys. Rev. A 59, 3336 (1999).
\bibitem{gt}D. Gottesman and I. L. Chuang, Nature (London) 402, 390 (1999);
M. A. Nielsen and I. L. Chuang, Quantum Computation and Quantum Information (Cambridge University Press, Cambridge, England, 2000).
\bibitem{duer}W. Duer, G. Vidal, and J.I. Cirac, Phys. Rev. A 62, 062314 (2000).
\bibitem{wtele}Bao-Sen Shi and Akihisa Tomita Physics Letters A
296,161(2002)
\bibitem{wbell}A. Cabello, Phys. Rev. A 65, 032108 (2002).
\bibitem{bells}A. Cabello, quant-ph/0205183.
\bibitem{hou}G. Teklemarian et al, Phys. Rev. A66, 012309(2002).
\bibitem{daa}D. Bouwmeester, A. Ekert and A. Zeilinger, The physics of Quantum Information( Springer-Verlag, Berlin, 2000)
\bibitem{db}D. Bouwmeester et al , Nature(London) {\bf 390} 575 (1997)
\bibitem{km}K. Mattle et al,  Phys. Rev.
Lett. {\bf 76}, 4656 (1996).
\bibitem{ds} D. S. Naik et al  Phys. Rev. Lett.  {\bf 84},
4733 (2000); W. Tittel et al  Phys. Rev. Lett.  {\bf 84}, 4737
(2000).
\bibitem{ggg}A. Zeilinger, M. A. Horne, H. Weinfurter, and M. Zukowski, Phys. Rev. Lett. 78, 3031 (1997).
\bibitem{bou}D. Bouwmeester , J-W. Pan, M. Daniell, H. Weinfurter , and A.
Zeilinger , Phys. Rev. Lett. 82, 1345 (1999).
\bibitem{dj}J. W. Pan et al, Phys. Rev. Lett.  {\bf 86}, 4435
(2001).
\bibitem{pan}J-W. Pan , D. Bouwmeester, M. Danlell, H. Weinfurter , and A. Zeilinger , Nature (London) 403, 515 (2000).
\bibitem{erg}E. Knill, R. Laflamme and G. Milburn, Nature(London) {\bf
409}, 46 (2001), T.C.Ralph, A.G.White, W.J.Munro, G.J.Milburn,
quant-ph/0108049;Terry Rudolph, Jian-Wei Pan,quant-ph/0108056;
XuBo Zou, K. Pahlke, W. Mathis, Phys. Rev. A 65, 064305 (2002)
\bibitem{zxb} Hwang Lee, Pieter Kok, Nicolas J. Cerf, Jonathan P.
Dowling,Phys. Rev. A 65,030101 (2002); Jaromir Fiurasek, Phys.
Rev. A 65, 053818 (2002); XuBo Zou, K. Pahlke and W. Mathis,
quant-ph/0110149, Phys. Rev. A 66, 014102 (2002)
\bibitem{gp}N. Gisin and A. Peres, Phys. lett. A 162, 15(1992)
\bibitem{kp}D. Kaszlikowski, P. Gnascinski, M. Zukowski, W. Miklaszewski, and A. Zeilinger, Phys. Rev. Lett. 85, 4418 (2000).
\bibitem{zxb1}XuBo Zou, K. Pahlke, W. Mathis, quant-ph/0111014
\bibitem{pdd}P. D. Drummond, Phys. Rev. Lett. 50, 407 (1983).
\bibitem{pdd1}A. Lamas-Linares
et. al. NATURE (London) 412, 887 (2001) J. C. Howell et al. PRL 88
030401 (2002)
\bibitem{kkk}P. G. Kwiat, K. Mattle, H. Weinfurter, A. Zeilinger, A. V. Sergienko, and Y. H. Shih, Phys. Rev. Lett. 75, 4337 (1995)
\bibitem{mandel} C. K. Hong, Z. Y. Ou, L. Mandel, Phys. Rev. Lett.
59, 2044 (1987)
\end{thebibliography}
\end{document}